\documentclass[lettersize,journal]{IEEEtran}
\usepackage{amsmath,amsfonts}
\usepackage{color}
\usepackage{array}
\usepackage{multirow}

\usepackage[caption=false,font=normalsize,labelfont=sf,textfont=sf]{subfig}
\usepackage{textcomp}
\usepackage{stfloats}
\usepackage{url}
\usepackage{wrapfig}
\usepackage{caption} %
\usepackage{helvet}  
\usepackage{courier}  
\usepackage{graphicx} %
\usepackage{hyperref}
\usepackage{algorithm}
\usepackage{algpseudocode} 
\usepackage{booktabs}
\usepackage[table]{xcolor}
\usepackage{amsmath}
\usepackage{hyperref}
\usepackage{utfsym}
\usepackage{amssymb}
\usepackage{multirow}
\usepackage{ulem}
\usepackage{amssymb}
\usepackage{verbatim}
\usepackage{newfloat}
\usepackage{listings}
\usepackage{graphicx}
\def\BibTeX{{\rm B\kern-.05em{\sc i\kern-.025em b}\kern-.08em
    T\kern-.1667em\lower.7ex\hbox{E}\kern-.125emX}}
\usepackage{balance}

\begin{document}
\title{\includegraphics[width=1.7cm,height=0.55cm]{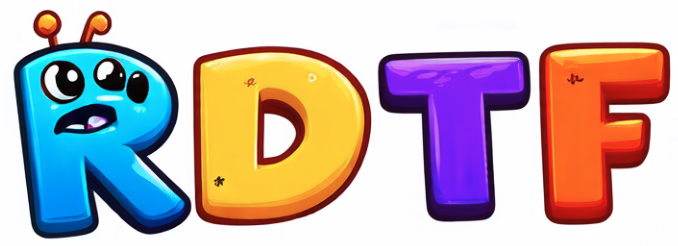} :\huge{Resource-Efficient Dual-Mask Training Framework for Multi-Frame Animated Sticker Generation}}

\author{Zhiqiang~Yuan,
        Ting~Zhang,
        Peixiang~Luo,
        Ying~Deng,
        Jiapei~Zhang, \\
        Zexi~Jia,
        Jinchao~Zhang*,
        Jie~Zhou ~\IEEEmembership{Senior Member,~IEEE}
    
\thanks{
\textit{Corresponding author is Jinchao Zhang (dayerzhang@tencent.com).}}
\noindent
\thanks{Z. Yuan, P. Luo, Y. Deng, J. Zhang, J. Zhou, and J. Zhang are with the Pattern Recognition Center, WeChat AI, Tencent Inc, Beijing 100193, China.
T. Zhang is with the University of Chinese Academy of Sciences, Beijing 100190, China.}
}

\markboth{Submitted to IEEE Transactions on Multimedia}%
{How to Use the IEEEtran \LaTeX \ Templates}

\maketitle

\begin{abstract}

Recently, significant advancements have been achieved in video generation technology, but applying it to resource-constrained downstream tasks like multi-frame animated sticker generation (ASG)—characterized by low frame rates, abstract semantics, and long-tail frame length distribution—remains challenging. Parameter-efficient fine-tuning (PEFT) techniques (e.g., Adapter, LoRA) for large pre-trained models suffer from insufficient fitting ability and source-domain knowledge interference.
In this paper, we propose Resource-Efficient Dual-Mask Training Framework (RDTF), a dedicated solution for multi-frame ASG task under resource constraints. We argue that training a compact model from scratch with million-level samples outperforms PEFT on large models, with RDTF realizing this via three core designs: 1) a Discrete Frame Generation Network (DFGN) optimized for low-frame-rate ASG, ensuring parameter efficiency; 2) a dual-mask based data utilization strategy to enhance the availability and diversity of limited data; 3) a difficulty-adaptive curriculum learning method that decomposes sample entropy into static and adaptive components, enabling ``easy-to-difficult'' training convergence.
To provide high-quality data support for RDTF’s training from scratch, we construct VSD2M—a million-level multi-modal animated sticker dataset with rich annotations (static and animated stickers, action-focused text descriptions)—filling the gap of dedicated animated data for ASG task.
Experiments demonstrate that RDTF is quantitatively and qualitatively superior to state-of-the-art PEFT methods (e.g., I2V-Adapter, SimDA) on ASG tasks, verifying the feasibility of our framework under resource constraints. Corresponding files are released at \textcolor{blue}{\href{https://mails9523.github.io}{link}}.

\end{abstract}

\begin{IEEEkeywords}
VSD2M, Vision-language Sticker Dataset, Multi-frame Animated Sticker Generation.
\end{IEEEkeywords}

\section{Introduction}

Video generation, driven by the rapid advancement of pre-trained diffusion models \cite{11153880}\cite{11153996}\cite{10589534}, has achieved breakthroughs in generating high-quality dynamic content, spanning natural scene synthesis, multimedia storytelling, and social interaction applications. 
Particularly in social communication scenarios, \textbf{animated sticker generation (ASG)}, a typical downstream video generation task characterized by low frame rates,  abstract semantic expression, and a long-tail distribution of frame lengths—has become a key medium for enhancing user interaction. 
However, translating state-of-the-art video generation capabilities into such specialized, resource-constrained downstream tasks remains a critical challenge.

\begin{figure}[!t]
\centering
\includegraphics [width=3.4in]{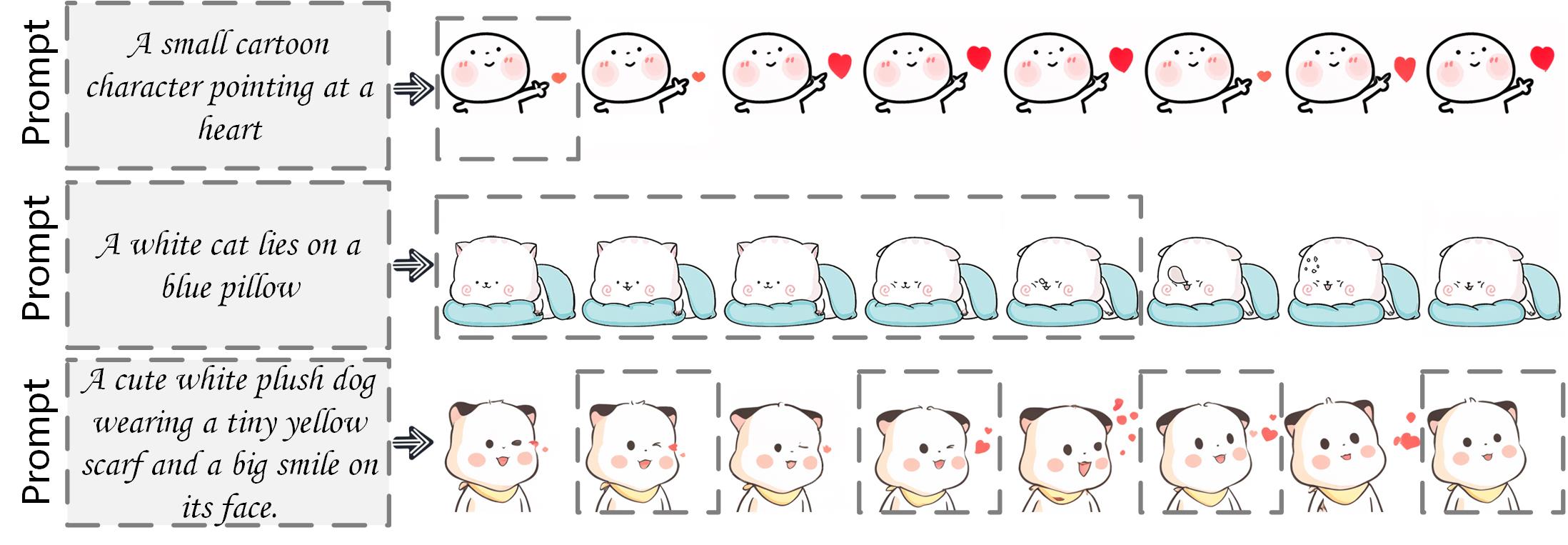}
\captionsetup{font={small},skip=4pt, justification=raggedright}
\caption{Results of Animated Sticker Generation via RDTF.
Our approach demonstrates excellent performance across diverse ASG tasks.
Gray boxes signify either textual or visual guidance.
See \textcolor{blue}{\href{https://mails9523.github.io}{here}} for dynamic results.
}
\vspace{-20px}
\label{plt_bar_framesnum}
\end{figure}

Pretrained large models (e.g., I2VGen-XL with 1.8B parameters) excel at general video generation but face significant barriers in resource-limited situations. 
For instance, full-parameter fine-tuning of I2VGen-XL is infeasible on a V100 GPU with 32GB memory for task-specific adaptation \cite{zhang2023i2vgen}. 
To address this, parameter-efficient fine-tuning (PEFT) techniques, such as Adapter \cite{10233038} and LoRA \cite{hulora}, have been widely adopted, as they only update a small subset of parameters (e.g., 24M parameters for SimDA \cite{xing2024simda} on a 1.1B-parameter base model). 
While PEFT reduces resource consumption, it suffers from two inherent limitations that are particularly pronounced in ASG-like specialized tasks:  
\textbf{a. Insufficient fitting ability}: Only a tiny fraction of parameters participate in gradient propagation, leading to weak adaptation to task-specific patterns.  
\textbf{b. Source-domain knowledge interference}: The fine-tuned model inherits dominant domain knowledge from pre-training (e.g., natural video motion), causing inference to deviate from the target domain.  
These limitations result in suboptimal performance for PEFT methods in downstream specialized tasks, highlighting the need for a new training paradigm that balances resource efficiency and task adaptation.

In this paper, we argue that under constrained computing resources, \textbf{training a compact video generation model from scratch using million-level samples} can outperform PEFT on large models for downstream tasks. The core of this paradigm lies in two pillars: \textit{task-aware data utilization} (to fully exploit limited samples) and \textit{adaptive curriculum learning} (to optimize the model’s learning route). This approach avoids the fitting bottlenecks and domain interference of PEFT, enabling the small model to learn robust, task-specific generative patterns—especially for specialized tasks like ASG.

To validate this paradigm, we take ASG as a case study and propose a \textbf{Resource-efficient Dual-mask Training Framework (RDTF)}, tailored to address the unique challenges of ASG task with resource constraints. RDTF consists of three interdependent components:  
1. \textbf{Discrete Frame Generation Network (DFGN)}: A compact noise prediction network designed for low-frame-rate ASG task. It integrates a Spatial-Temporal Interaction (STI) layer that splits temporal modeling into semantic interaction (cross-frame attention for global motion) and detail maintenance (1×1 convolutions for fine-grained texture preservation), ensuring parameter efficiency while capturing sticker-specific frame discreteness.  
2. \textbf{Dual-mask based Data Utilization (DDU) strategy}: To maximize the value of limited ASG data, a condition mask enables multi-task joint training (interpolation, pre/post-frame prediction, text-image guided generation), while a loss mask—coupled with feature clustering—addresses the long-tail distribution of sticker frame lengths, improving data information density and diversity.  
3. \textbf{Difficulty-adaptive Curriculum Learning (DCL)}: To facilitate smooth convergence under dual-mask training, DCL decomposes sample entropy into a static component (ensuring global entropy increase by adjusting task probabilities) and an adaptive component (dynamically tuning difficulty via PID-based historical loss analysis), ensuring the model learns from easy to difficult samples.  

To provide sufficient data support for training the compact model from scratch, we construct \textbf{VSD2M}, a million-level vision-language dataset specialized for ASG task to address the scarcity of high-quality animated sticker datasets. VSD2M contains 2.09M samples (including static and animated stickers, and paired text descriptions) with rich annotations, filling the gap of dynamic, task-aligned data in existing sticker datasets (e.g., SER30K \cite{liu2022ser30k}, Sticker820K \cite{zhao2023sticker820k}) which are limited to static content.

The main contributions of this work are summarized as follows:  
\begin{itemize}
    \item We validate a new training paradigm: Under constrained resources, a compact video generation model trained from scratch with million-level samples outperforms PEFT on large models in downstream tasks.  
    \item We propose RDTF, a resource-efficient framework for video generation that integrates DFGN (task-aware network design), DDU (dual-mask data utilization), and DCL (adaptive curriculum learning) to achieve superior performance under resource constraints. 
    \item We construct VSD2M, a large-scale multi-modal animated sticker dataset with 2.09M samples and rich annotations, providing critical data support for training compact ASG models from scratch and filling the gap of dynamic sticker data.
    \item Our ASG-specialized method based on RDTF achieves state-of-the-art performance on VSD2M, outperforming I2V-Adapter, SimDA, and Customize-A-Video in FVD, VQA, and CLIP similarity.
\end{itemize}


\section{Related Works}

\subsection{Video Generation Model} 

Early video generation methods primarily relied on Generative Adversarial Networks (GANs) \cite{creswell2018generative}, such as Temporal GAN \cite{saito2017temporal} and MoCoGAN \cite{tulyakov2018mocogan}, which achieved temporal consistency by learning the joint distribution of video frames. However, GAN-based methods often struggled with mode collapse and limited scalability for long sequences. The above limitations partially addressed by Transformer-based models (e.g., VideoGPT \cite{yan2021videogpt}, CoGVideo \cite{hong2022cogvideo}) that leveraged self-attention \cite{vaswani2017attention} to model long-range temporal dependencies, improving efficiency but remaining computationally heavy for resource-constrained scenarios.

In recent years, Diffusion Probabilistic Models (DPMs) have become dominant in high-quality video generation. Works like Imagen Video \cite{ho2022imagen} and VideoFusion \cite{luo2023videofusion} optimized DPM architectures by integrating temporal modules into 2D U-Nets, while Align Your Latents \cite{blattmann2023align} focused on high-resolution synthesis. DiT-based models (e.g., Sora \cite{liu2024sora}, Open-Sora \cite{opensora}) further enhanced capabilities via scaling, but their large parameter sizes (often billions) make fine-tuning infeasible on limited hardware (e.g., V100 32GB GPU). Critically, these methods are designed for general natural videos (e.g., continuous motion, high frame rates) and fail to address the unique characteristics of animated sticker generation — low frame rates, abstract semantics, and long-tail frame length distribution — leaving a gap for task-specialized, resource-efficient models.

\vspace{-10px}
\subsection{Parameter-efficient Tuning for Video Generation}

To adapt large pre-trained models to downstream tasks under constraints, Parameter-Efficient Fine-Tuning techniques have emerged. AnimateDiff \cite{guo2023animatediff} proposed MotionLoRA to adapt pre-trained motion modules with low data costs; I2V-Adapter \cite{guo2024i2v} retained T2I model integrity while linking input conditions to self-attention via lightweight adapters; SimDA \cite{xing2024simda} fine-tuned only 24M parameters (out of 1.1B in the base T2I model) to enable video generation.

While PEFT reduces resource consumption, it suffers from two inherent flaws highlighted in our work: 1) insufficient fitting ability: limited trainable parameters (e.g., 24M in SimDA) fail to capture task-specific patterns like ASG’s discrete frame dynamics; 2) source-domain interference: inherited pre-training knowledge (e.g., natural video motion) deviates inference from the target domain. Even with enhancements (e.g., applying our dual-mask data utilization (DDU) to I2V-Adapter), PEFT cannot fully overcome these limitations—creating a need for a paradigm shift away from large-model tuning.

\vspace{-10px}
\subsection{Curriculum Learning}

Inspired by human learning, curriculum learning \cite{bengio2009curriculum, hacohen2019power} trains models from simple to complex tasks. In diffusion models, this has been applied to sequence data types (e.g., Any-to-Any Generation \cite{tang2024any}) or noise level discretization (e.g., consistency models \cite{song2023improved}), which gradually increase noise discretization steps to stabilize training.

However, existing curriculum strategies lack adaptability to multi-task, dual-mask training scenarios (e.g., ASG's interpolation, prediction, and generation tasks). They either rely on fixed global difficulty schedules (e.g., linear noise adjustment \cite{song2023improved}) or task-specific prior knowledge \cite{kim2024denoising}, failing to address the entropy instability caused by independent mask controls (frame length vs. task type). Our difficulty-adaptive curriculum learning (DCL) fills this gap by decomposing entropy into static (global increasing trend) and adaptive (PID-based historical loss adjustment) components—enabling stable convergence under dual-mask training.

\subsection{Sticker Vision-Language Datasets}

Unlike common natural-scene Visual-Language Datasets (VLD) \cite{poria2018meld, fei2021towards}, sticker-oriented VLD remains scarce despite stickers being a common human interactive tool. To address the scarcity of sticker data, Liu \textit{et al.} \cite{liu2022ser30k} built SER30K—a large-scale sticker emotion recognition dataset with 30K static stickers covering 1887 topics and emotion labels. Zhao \textit{et al.} \cite{zhao2023sticker820k} further collected Sticker820K, an 820K sample sticker-text dataset for retrieval and captioning tasks. However, these datasets focus on static stickers, lacking the dynamic motion of animated ones. Thus, we built a million-scale multi-frame animated sticker dataset to advance multi-modal sticker-oriented visual-language generation.

\begin{figure*}[!t]
\centering
\includegraphics [width=6.0in]{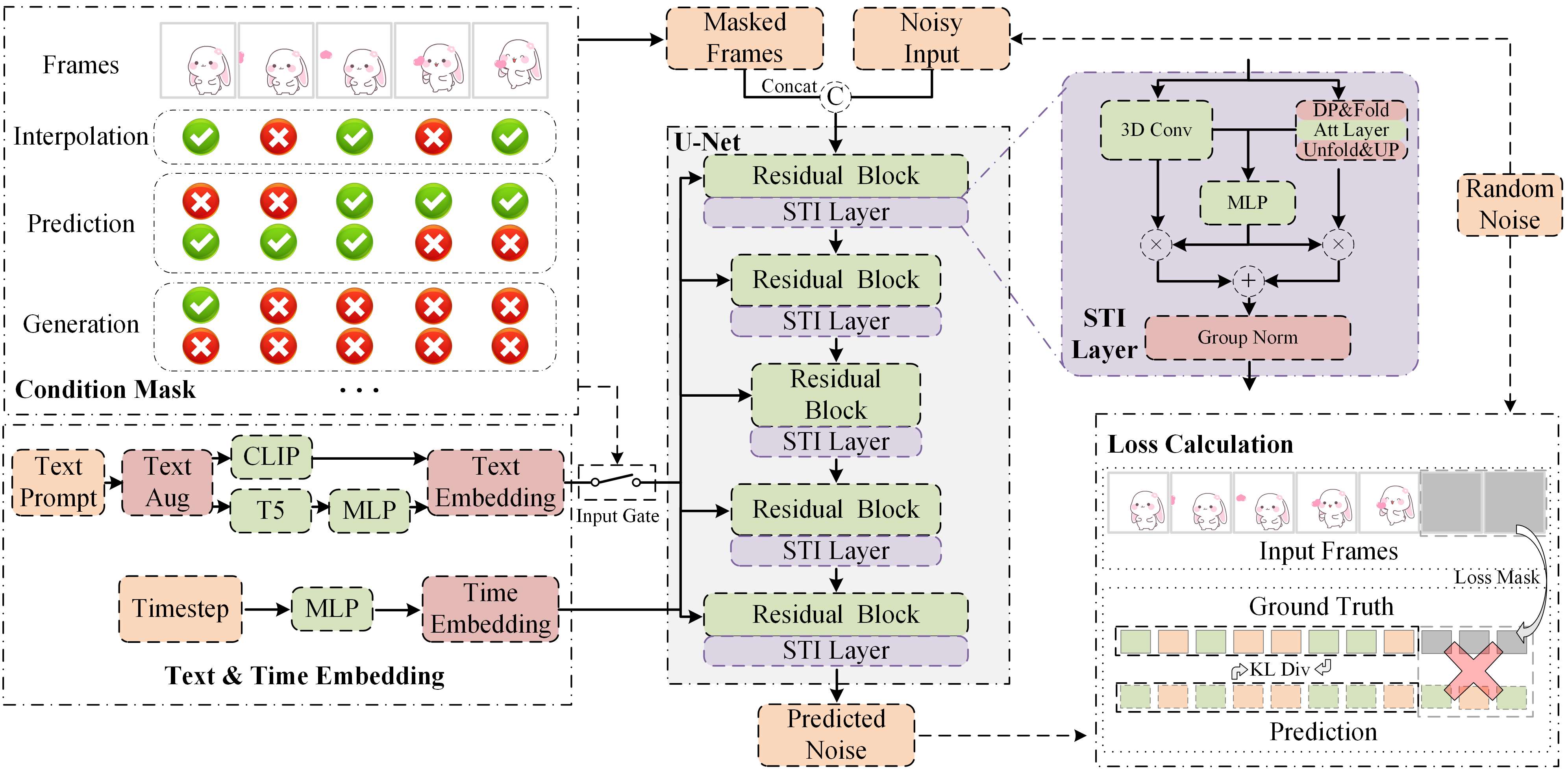}
\captionsetup{font={small},skip=4pt, justification=raggedright}
\caption{
Overview of resource-efficient dual-mask training framework.
We propose a discrete frame generation network to model the discreteness between animated sticker frames.
Furthermore, the dual masks, $i.e.$, condition mask and loss mask, are designed to improve the availability and expand the diversity of limited data.
The difficulty-adaptive curriculum learning is applied to facilitate convergence.
}
\vspace{-13px}
\label{overview}
\end{figure*}

\section{Method}

\subsection{Definitions}

Assuming a paired dataset with samples $(\boldsymbol{x},\boldsymbol{y}) \sim p_{\text{train}}(\boldsymbol{x},\boldsymbol{y})$, where $\boldsymbol{x}$ denotes the video clip with $N$ frames and $\boldsymbol{x}=\{x^i | i=1,2,...,N\}$, $\boldsymbol{y}$ is the corresponding caption.
DPMs \cite{song2020score} aims to sample from the distribution $p(x^i)$ by denoising and converting samples from a Gaussian distribution into target distribution iteratively, which can be divided into diffusion process and reverse process.
In step $t$ during diffusion process, a noised image $x^i_t$ is generated by $x^i_t = \sqrt{\alpha_t} x^i + \sqrt{(1-\alpha_t)} \epsilon$, $\epsilon \sim \mathcal{N}(0, \boldsymbol{I}_d)$, where $\epsilon$ is a Gaussian noise and $\alpha_t$ controls the noise ratio at step $t$.
In the reverse process, a learnable network $\mathcal{G}_{\theta_1}(x^i_t, t)$ aims to predict the noise and recover the clean image from the noised input $x^i_t$.
After training, the model starts from a pure Gaussian noise $x_T \sim \mathcal{N}(0, \boldsymbol{I}_d)$ and samples a clean image by iteratively running for $T$ steps.

Conditional video diffusion model \cite{batzolis2021conditional} regards the caption $\boldsymbol{y}$ and image $\boldsymbol{z}$ as guidance information to generate the paired video clip $\boldsymbol{x}$.
In this case, the model samples from the conditional distribution of $p(\boldsymbol{x}| \boldsymbol{y}, \boldsymbol{z})$, and the learnable network $\mathcal{G}_{\theta_2}
([\boldsymbol{x}_t, \boldsymbol{y},\boldsymbol{z}], t)$ parameterized by $\theta_{2}$ utilize the $\boldsymbol{x}_t$, $\boldsymbol{y}$ and $\boldsymbol{z}$ as input.
During training, the loss function can be modeled as:
\begin{equation}
    \mathcal{L}_t=
     \lVert \mathcal{G}_{\theta_2}( \boldsymbol{y}, \sqrt{\alpha_t} \boldsymbol{x}+ \sqrt{1-\alpha_t} \epsilon_t, \alpha_t ) - \epsilon_t \rVert^2
\label{eqa:diffusion-loss}
\end{equation}
where $\mathcal{L}_t$ is the loss function and $\epsilon_t$ is the noise at step $t$.

\vspace{-10px}
\subsection{Discrete Frame Generation Network}
To address the unique characteristics of animated sticker generation — low frame rates and abstract semantics — we propose the \textbf{Discrete Frame Generation Network (DFGN)} (Fig. \ref{overview}), a compact architecture built on the U-Net backbone \cite{unet} with task-specific optimizations for resource efficiency and sticker detail preservation. Its core innovation lies in the Spatial-Temporal Interaction (STI) layer and tailored multi-modal guidance fusion, as detailed below.

The STI layer is designed to model inter-frame dependencies in low-frame-rate stickers while avoiding redundant computations, splitting temporal modeling into two complementary branches:
\textbf{1. Semantic Interaction (Global Motion Modeling).}
To capture cross-frame semantic correlations (e.g., a character’s gesture change across sticker frames), we first process the input feature tensor \( \boldsymbol{h} \in \mathbb{R}^{F \times H \times W \times d} \) (where \( F \) is frame count, \( H/W \) is spatial resolution, \( d \) is feature dimension). We downsample \( \boldsymbol{h} \) by \( \gamma \) (to reduce self-attention complexity) to get \( \boldsymbol{h}_\gamma \in \mathbb{R}^{F \times H/\gamma \times W/\gamma \times d} \), then unfold its temporal-spatial dimensions into \( \mathbb{R}^{F \cdot H/\gamma \cdot W/\gamma \times d} \). Self-attention is applied to this flattened tensor to enable interaction between regions across frames; finally, \( \boldsymbol{h}_\gamma \) is upsampled by \( \gamma \) and reshaped to recover the original feature size \( F \times H \times W \times d \).
\textbf{2. Detail Maintenance (Local Texture Preservation).}
Semantic interaction may cause loss of fine-grained sticker details (e.g., text, lines, and color blocks). To mitigate this, we use convolutions with kernel size \( j \times 1 \times 1 \) —these operate along the channel dimension only, preserving spatial details while supplementing local feature information. After extracting semantic and detail features, we normalize both branches and fuse them via a linear layer, enabling dynamic feature integration across different network depths.

DFGN integrates image and text guidance to ensure ASG content consistency and style controllability, with designs optimized for sticker generation:
\textbf{Image Guidance}: Visual reference frames (e.g., static sticker sketches) are downsampled to match the VAE encoder’s latent output size (consistent with our training setup), then concatenated directly with noisy frames and fed into U-Net. Unlike prior works \cite{zhang2023i2vgen, voleti2024sv3d} that use cross-attention for visual injection, this concatenation avoids interfering with text controllability, which is critical when paired with the subsequent dual-mask strategy.
\textbf{Text Guidance}: To enhance generalization to diverse sticker captions, we randomly replace 10\% of text tokens (i.e., \( p=0.1 \)) with the symbol “\#” for data augmentation. CLIP and T5 encoders extract text features from the augmented captions; these features are mapped to the model’s feature dimension via linear layers and input as text embeddings.

\vspace{-10px}
\subsection{Dual-mask based Data Utilization Strategy}

To address the two key challenges of animated sticker generation under limited data: \textbf{insufficient data diversity} and \textbf{long-tail distribution of frame lengths}, we propose a dual-mask based data utilization (DDU) strategy. It consists of a \textbf{condition mask} and a \textbf{loss mask}, which synergistically improve data availability and expand pattern diversity,  to provide critical data support for training compact models from scratch.

\begin{figure}[htbp]
\centering
\includegraphics [width=3.0in]{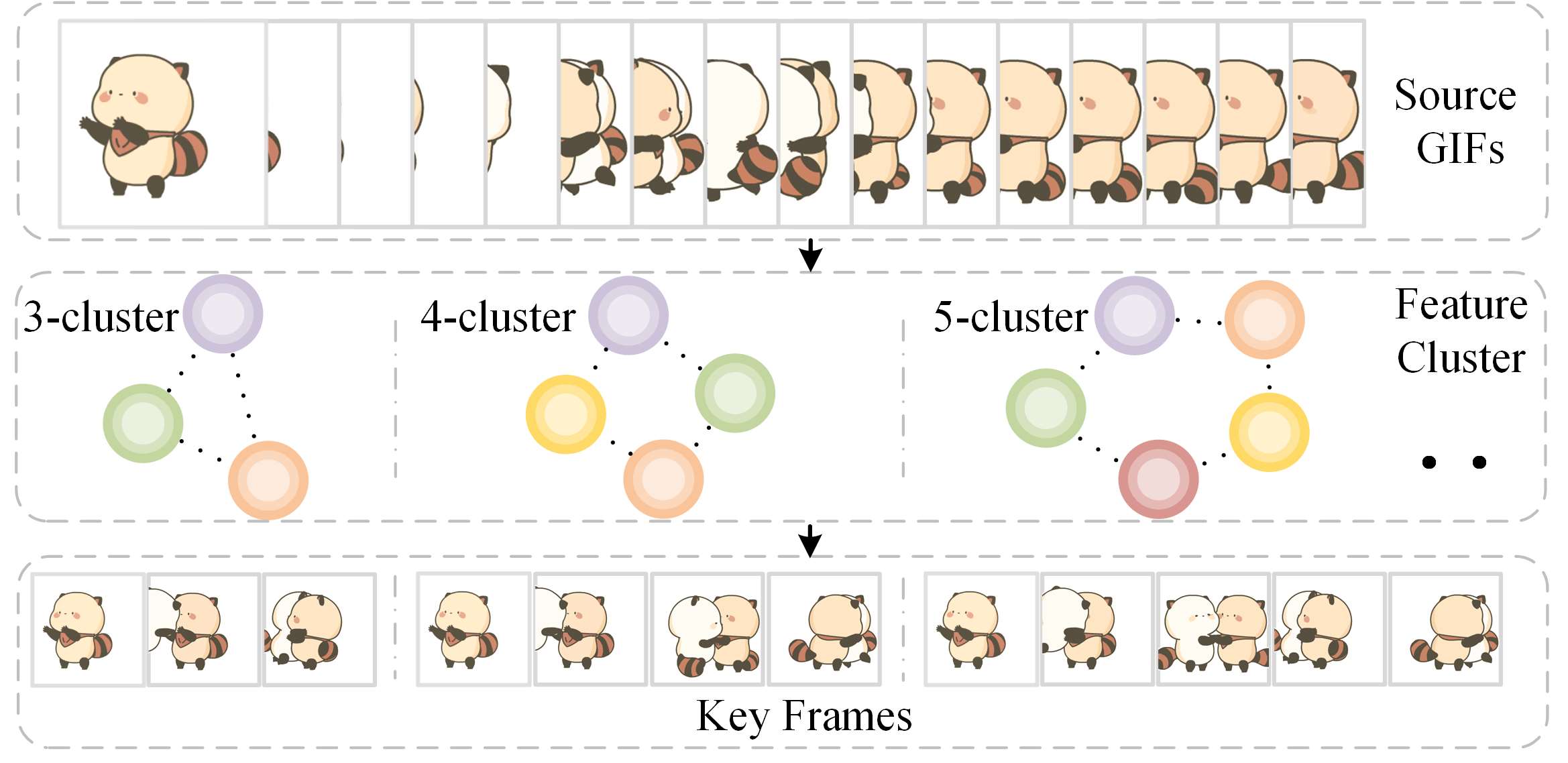}
\captionsetup{font={small},skip=4pt, justification=raggedright}
\caption{Frame extraction algorithm based on feature clustering. During training, data are clustered into $k$ clusters randomly to increase the information density.
}
\vspace{-10px}
\label{cluster}
\end{figure}

\noindent
\textbf{Condition Mask for Multi-task Data Reuse.}
The condition mask reuses limited ASG data by dynamically masking partial image/text guidance, enabling simultaneous training of three complementary tasks (Fig. \ref{overview}) that cover core ASG scenarios:
\textbf{Interpolation (IPT)}: Mask out intermediate frames of a sticker clip and train the model complete missing frames, enhancing its ability to model smooth inter-frame transitions.
\textbf{Pre \& Post Prediction (PDT)}: Mask out leading or trailing frames  and train the model to predict the masked part, strengthening its understanding of temporal continuity.
\textbf{Text \& Image-based Generation (GRT)}: Mask out most frames  and guide generation via text $\boldsymbol{y}$ and reference image $\boldsymbol{z}$, aligning with real-world ASG requirements.
While the condition mask shares the idea of ``masked guidance'' with MCVD \cite{voleti2022mcvd}, their core goals differ: MCVD uses masking to define separate tasks, whereas our condition mask is a data utilization strategy. It reuses the same dataset across tasks to maximize sample efficiency, with implementation details (e.g., mask pattern randomness, task weight balancing) tailored to ASG task’s low-frame-rate characteristics.

\begin{figure}[!htbp]
\centering
\includegraphics [width=3.0in]{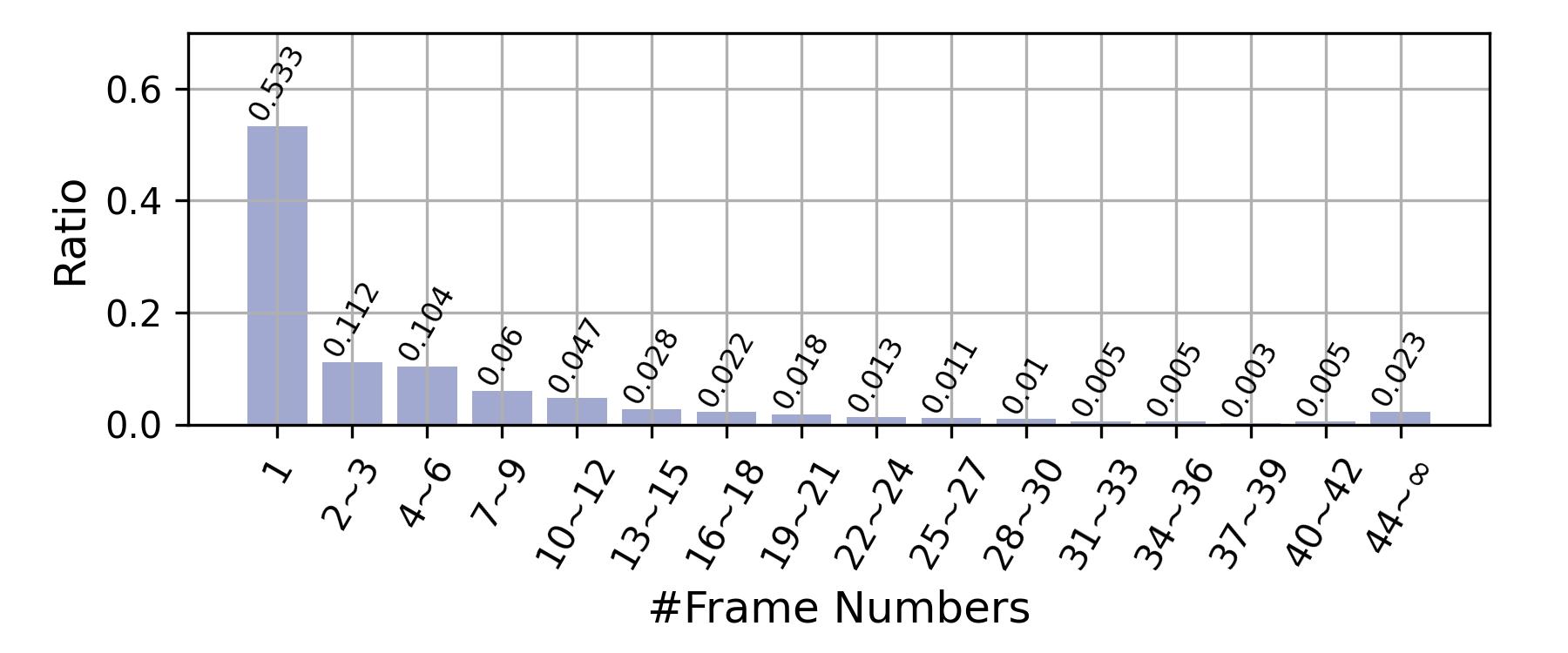}
\captionsetup{font={small},skip=4pt, justification=raggedright}
\caption{Frame distribution in collected sticker dataset, which follows the long-tail distribution, $i. e.$, more short frames and fewer long frames.
It poses challenges for fully utilizing long-frame data, highlighting the necessity of our clustering-based frame extraction and dual-mask strategy to enhance data utilization efficiency.
}
\vspace{-5px}
\label{plt_bar_framesnum}
\end{figure}

\noindent
\textbf{Loss Mask for Long-tail Data Density Enhancement.}
ASG data exhibits a long-tailed frame length distribution: most segments have short frames, while long frame segments containing richer motion patterns are rare.
Traditional equal-interval frame extraction wastes long-frame data and fails to exploit their information. To address this, we first apply feature clustering to long-frame clips during training: for a clip with $N>k$ frames (where $k \in [3,8]$ is randomly selected per batch), we cluster its frame features into $k$ clusters using Euclidean distance, ensuring each cluster retains distinct motion patterns. Then, a loss mask that restricts loss calculation is applied to only the first $k$ clustered frames (discarding redundant frames in the same cluster). This not only reduces computational overhead but also increases the information density of long-frame data. Each cluster subset exposes a unique pattern to the model, effectively expanding the diversity of limited training samples.

The condition mask (multi-task reuse) and loss mask (long-tail density enhancement) form a closed loop: the condition mask maximizes the use of each sample across tasks, while the loss mask unlocks latent patterns in underutilized long-frame data. Together, they transform ASG training samples into a more diverse, information-dense dataset, so as to solve the ``data scarcity'' problem for training compact models from scratch, and laying the foundation for stable convergence of the subsequent curriculum learning strategy.

\vspace{-10px}
\subsection{Difficulty-adaptive Curriculum Learning}
\begin{figure}[!h]{}
  \begin{center}
  \vspace{-10px}
  \includegraphics[width=0.21\textwidth]{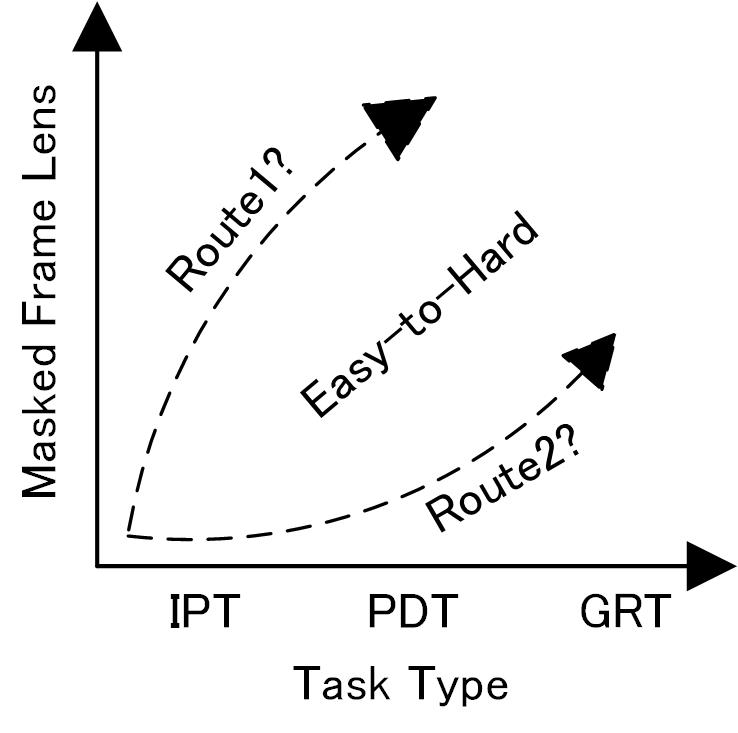}
  \end{center}
  \vspace{-15px}
  \caption{The masked frame length and task type during training are independent of each other, which makes it difficult to determine a route to obtain entropy-increasing samples stably.}
  \label{CLproblem}
  \vspace{-10px}
\end{figure}

In resource-constrained animated sticker generation training, a critical challenge arises: as shown in Fig. \ref{CLproblem}, the independence of masked frame length ($N$) and task type ($T$) breaks the monotonicity of sample entropy $\mathcal{H}_t$, which is required for stable curriculum learning \cite{bengio2009curriculum}. To address this, we propose a Difficulty-adaptive Curriculum Learning (DCL) strategy, which decomposes $\mathcal{H}_t$ into complementary components via entropy additivity and ensures $\mathcal{H}_t > \mathcal{H}_{t-1}$ for all training steps $t$, facilitating convergence of models trained from scratch.

\noindent
\subsubsection{\textbf{Entropy Instability in Dual-mask Training}}

For ASG task, sample entropy $\mathcal{H}_t$ is determined by two below orthogonal factors.  
\textit{a. Task type entropy}: $\mathcal{H}^{(IPT)} < \mathcal{H}^{(PDT)} < \mathcal{H}^{(GRT)}$, where $IPT$ (interpolation), $PDT$ (pre/post prediction), and $GRT$ (generation) represent increasing guidance sparsity (thus higher uncertainty).  
\textit{b. Frame length entropy}: $\mathcal{H}^{(T,N-1)} < \mathcal{H}^{(T,N)}$ for any task $T$, as longer frames introduce more temporal dependencies.  

Due to the independence of $T$ and $N$, naive curriculum learning cannot guarantee $\mathcal{H}_t > \mathcal{H}_{t-1}$. For example, a sample with $(GRT, N=3)$ (high task entropy, low frame entropy) may have lower $\mathcal{H}_t$ than $(PDT, N=8)$ (low task entropy, high frame entropy), leading to random difficulty fluctuations. To solve this, we leverage the \textbf{additivity of entropy} to split $\mathcal{H}_t$ into two controllable components.

\noindent
\subsubsection{\textbf{Entropy Decomposition}}

By the additivity of entropy, sample entropy $\mathcal{H}_t$ can be decomposed into a \textit{static component $\overline{\mathcal{H}}_t$} (for global difficulty trend) and an \textit{adaptive component $\tilde{\mathcal{H}}_t$} (for local loss correction), i.e.:  
\begin{equation}
\label{eq1}
\mathcal{H}_t = \lambda \cdot \overline{\mathcal{H}}_t + (1-\lambda) \cdot \tilde{\mathcal{H}}_t
\end{equation}  
where $\lambda \in (0,1)$ is a balance coefficient that harmonizes global consistency and local adaptability. Below, we prove the monotonicity ($\overline{\mathcal{H}}_t > \overline{\mathcal{H}}_{t-1}$ and $\tilde{\mathcal{H}}_t \geq \tilde{\mathcal{H}}_{t-1}$) of each component.

\noindent
\textbf{Static Component $\overline{\mathcal{H}}_t$: Ensuring Global Entropy Growth.}
The static component enforces a global ``easy-to-difficult'' order by adjusting the static probability of task-type combinations ($\overline{\mathcal{P}}_t^{T,N}$), where $T \in \{IPT, PDT, GRT\}$ and $N$ is frame length.  
First, define the intrinsic entropy of task $T$ as $\mathcal{H}^T$ (satisfying $\mathcal{H}^{IPT} < \mathcal{H}^{PDT} < \mathcal{H}^{GRT}$) and the intrinsic entropy of frame length $N$ as $\mathcal{H}^N$ (satisfying $\mathcal{H}^{N-1} < \mathcal{H}^N$). The static entropy $\overline{\mathcal{H}}_t$ is the weighted sum of intrinsic entropies, with weights being static probabilities:  
\begin{equation}
\overline{\mathcal{H}}_t = \sum_{T,N} \overline{\mathcal{P}}_t^{T,N} \cdot (\mathcal{H}^T + \mathcal{H}^N)
\end{equation}  
To ensure $\overline{\mathcal{H}}_t > \overline{\mathcal{H}}_{t-1}$, we design $\overline{\mathcal{P}}_t^{T,N}$ to follow two rules:  
a. For fixed $N$: $\overline{\mathcal{P}}_t^{IPT,N} \leq \overline{\mathcal{P}}_{t-1}^{IPT,N}$ and $\overline{\mathcal{P}}_t^{PDT,N}, \overline{\mathcal{P}}_t^{GRT,N} \geq \overline{\mathcal{P}}_{t-1}^{PDT,N}, \overline{\mathcal{P}}_{t-1}^{GRT,N}$ (shift weight from low-entropy $IPT$ to high-entropy $PDT/GRT$).  
b. For fixed $T$: $\overline{\mathcal{P}}_t^{T,N} \geq \overline{\mathcal{P}}_{t-1}^{T,N}$ for increasing $N$ (shift weight from short frames to long frames).  
According to these rules, the weighted sum $\overline{\mathcal{H}}_t$ is monotonically increasing because high-entropy combinations $(T,N)$ acquire greater weight over time.
Therefore, the static component guarantees a global difficulty upward trend, even if $T$ and $N$ are independent.

\noindent
\textbf{Adaptive Component $\tilde{\mathcal{H}}_t$: Correcting Local Entropy Fluctuations.}
The static component cannot resolve local fluctuations (e.g., a $(PDT, N=8)$ sample may be harder than expected if the model’s loss spikes). The adaptive component $\tilde{\mathcal{H}}_t$ addresses this by linking difficulty to historical loss statistics, ensuring local entropy stability.  
First, define the loss deviation $\mathcal{L}_s(t)$ to measure the difference between current batch difficulty and historical average difficulty:  
\begin{equation}
\mathcal{L}_s(t) = \mathcal{L}_c(t) - \frac{1}{t} \int_{0}^{t} \mathcal{L}_c(\tau) d\tau
\end{equation}  
where $\mathcal{L}_c(t)$ is the current batch loss, and $\frac{1}{t} \int_{0}^{t} \mathcal{L}_c(\tau) d\tau$ is the average loss up to step $t$. A positive $\mathcal{L}_s(t)$ indicates the current batch is harder than average; a negative value indicates it is easier.  
To stabilize $\mathcal{L}_s(t)$ and map it to adaptive difficulty, we use a proportional-integral-derivative (PID) controller, which outputs a joint difficulty score $\tilde{\mathcal{P}}^*$:  
\begin{equation}
\label{pid}
\tilde{\mathcal{P}}^* = \mathcal{K}_p \cdot \mathcal{L}_s(t) + \mathcal{K}_i \cdot \int_{0}^{t} \mathcal{L}_s(\tau) d\tau + \mathcal{K}_d \cdot \frac{d}{dt}\mathcal{L}_s(t)
\end{equation}  
where $\mathcal{K}_p, \mathcal{K}_i, \mathcal{K}_d$ are non-negative control coefficients (controlling system stability, response speed, and steady-state error).  
The adaptive entropy $\tilde{\mathcal{H}}_t$ is a strictly increasing function of $\tilde{\mathcal{P}}^*$, i.e.:  
\begin{equation}
\tilde{\mathcal{H}}_t = g(\tilde{\mathcal{P}}^*)
\end{equation}  
where $g(\cdot)$ satisfies $g(a) > g(b)$ if $a > b$. When $\mathcal{L}_s(t) > 0$ (current batch is too hard), $\tilde{\mathcal{P}}^*$ decreases, lowering $\tilde{\mathcal{H}}_t$ to select easier samples; when $\mathcal{L}_s(t) < 0$ (current batch is too easy), $\tilde{\mathcal{P}}^*$ increases, raising $\tilde{\mathcal{H}}_t$. This ensures $\tilde{\mathcal{H}}_t$ adjusts dynamically to maintain local entropy growth, complementing the static component.

\begin{figure*}[!t]
\centering
\includegraphics [width=6.2in]{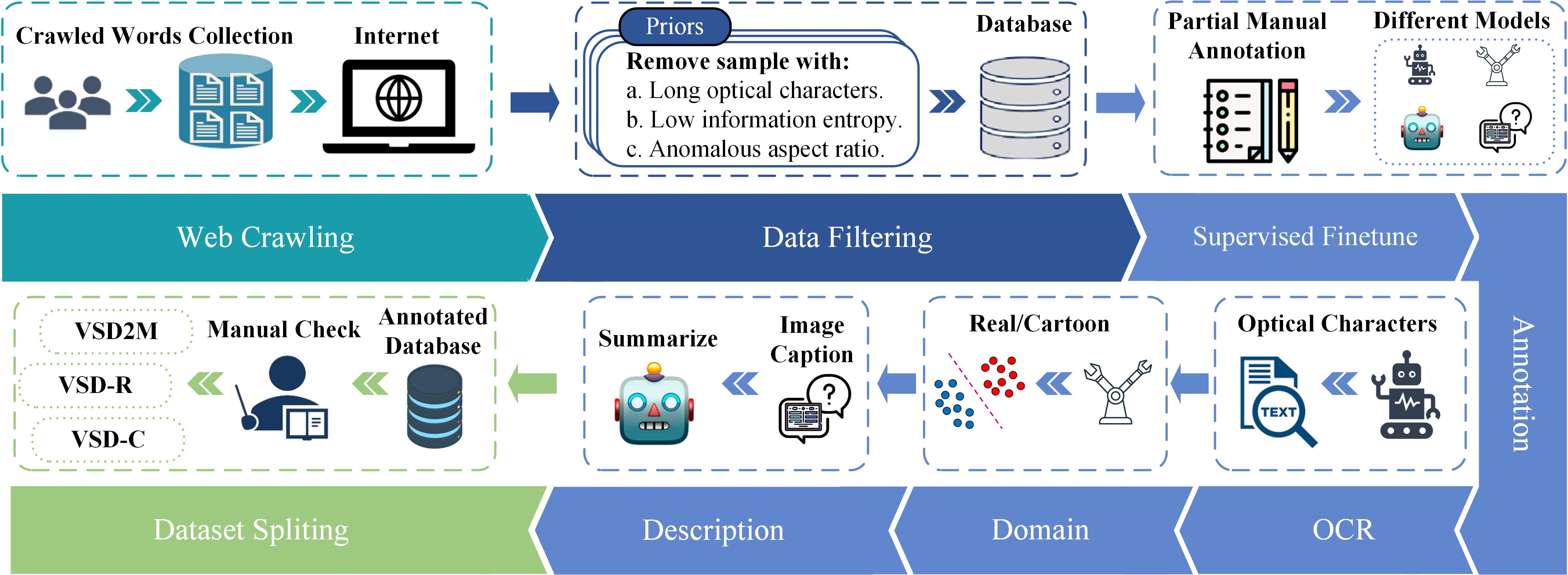}
\captionsetup{font={small},skip=4pt, justification=raggedright}
\caption{
Overview of data collection and processing, which can be divided into four stages: web crawling, data filtering, annotation and dataset splitting.
During the data annotation process, we use manually labeled data to fine-tune different models to obtain high-quality semi-automatic annotation results.
}
\label{pipeline}
\end{figure*}

\noindent
\subsubsection{\textbf{Component Fusion}}
From above subsections, we have:  
a. $\overline{\mathcal{H}}_t > \overline{\mathcal{H}}_{t-1}$ (global static trend, proven via weighted entropy growth);  
b. $\tilde{\mathcal{H}}_t \geq \tilde{\mathcal{H}}_{t-1}$ (local adaptive correction, guaranteed by PID-based difficulty adjustment).  
Substituting into the entropy decomposition formula (Eq. \ref{eq1}), the final sample entropy satisfies:  
\begin{equation}
\mathcal{H}_t = \lambda \overline{\mathcal{H}}_t + (1-\lambda) \tilde{\mathcal{H}}_t > \lambda \overline{\mathcal{H}}_{t-1} + (1-\lambda) \tilde{\mathcal{H}}_{t-1} = \mathcal{H}_{t-1}
\end{equation}  
This mathematically proves that DCL ensures monotonic sample entropy growth, so as to solve the core problem of dual-mask training instability in ASG task.

This section mathematically demonstrates DCL’s feasibility via entropy decomposition and control theory, addressing the dual-mask instability problem while adhering to the resource-constrained premise of ASG training.
The formula-driven design of DCL is validated by experiments:  
a. Quantitative results (Table \ref{ablation}): DCL outperforms linear curriculum learning (LCL) in Fréchet Video Distance (FVD) and Video Quality Assessment (VQA), as LCL lacks adaptive correction and fails to maintain entropy monotonicity.  
b. Qualitative convergence (Fig. \ref{loss}): DCL’s loss curve is smoother than LCL and SimDA, confirming that the static-adaptive fusion stabilizes training—consistent with our proof of $\mathcal{H}_t > \mathcal{H}_{t-1}$.

\begin{table*}[!t]
\centering
\resizebox{\linewidth}{!}{
\begin{tabular}{c|ccccccc}
\hline
Dataset     & Size & Modality &     language      & Trigger Word & OCR & Real/Cartoon Flag & Is Open \\ \hline
TGIF \cite{li2016tgif}  & 100K & GIFs, description & En & & & &  \checkmark \\
MOD \cite{fei2021towards}      & 45K    & static sticker &                      &        &      &     & \checkmark
\\
SRS \cite{gao2020learning}      & 320K    & static sticker &      &       &      &   & \checkmark             \\
SER30K \cite{liu2022ser30k}      & 30K    & static sticker & En                     & \checkmark       &      &        \checkmark     & \checkmark    \\
Sticker820K \cite{zhao2023sticker820k} & 820K   & static sticker, description & En     & \checkmark        & \checkmark   &        &       \\
VSD2M        & \textbf{2.09M}    & \textbf{static sticker, GIFs, description} & En, \textbf{Cn}  & \checkmark       & \checkmark     & \textbf{\checkmark}  & \checkmark              \\ \hline
\end{tabular}
}
\caption{
Static information comparison of different vision-language sticker datasets.
}
\label{datasets_comp}
\end{table*}

\section{VSD2M DATASET}
\label{mvsd}

\subsection{Data Construction}

In this subsection, we construct a million-level vision-language sticker dataset to facilitate the ASG task.

Fig.\ref{pipeline} show the overview of data collection and processing.
We initially obtained 2.5M data from Internet
by retrieving 600K crawled words, and then the samples that meet the following conditions were removed based on observed priors: 
\textit{
a. The sample with long optical characters.
b. The sample with low information entropy.
c. The sample with anomalous aspect ratio.
}
After this, the candidate data which contains 2.1M static and animated stickers are obtained.

To reduce the semantic gap between modalities, several different strategies are utilized to enrich the dataset labels.
We manually labeled portions of the data for different annotation tasks, allowing us to fine-tune the following models.
\textbf{a. OCR.}
PaddleOCR \cite{du2020pp} finetuned on the labeled sticker dataset is used to extract optical characters from visual samples.
\textbf{b. Domain.}
We train a classification model based on EfficientNet \cite{koonce2021efficientnet} to provide discrimination between real and cartoon scenes.
\textbf{c. Description.}
For GIFs, we use 330K pieces of manually annotated data to perform supervised fine-tuning on VideoLlama \cite{zhang2023video} to improve the accuracy of automatic annotation, so that the trained model can be directly used for visual descriptions generation.
Different from Sticker820K \cite{zhao2023sticker820k}, the description in VSD2M contains the action information of the GIFs, which has an important impact on sticker generation control \cite{yang2020captionnet}.
While for static stickers, we use LLaVa \cite{liu2023visual} to extract description.
After obtaining descriptions of static and animated stickers, we use a large language model to produce cleaner captions by summarizing the results, where the prompt leveraged is:
``\texttt{Summarize the following text from both static and dynamic features: \textbackslash n \textcolor{orange}{[Description]}}''.
In addition to the English version, we also provide a Chinese version to facilitate the needs of researchers.
\textbf{d. Trigger word.}
Crawled words related to the visual sample are aggregated and regarded as trigger words.
The comparison of static information between the VSD2M and other sticker-related datasets is shown in Table \ref{datasets_comp}.
VSD2M incorporates GIFs modality compared to other sticker datasets and is greatly ahead in terms of data scale.

\begin{table}[!t]
\resizebox{\linewidth}{!}{
\begin{tabular}{ccc}
\hline
Multi-frame Ratio    &  Optical Character Ratio   &   Ave Frames Numbers       \\ \hline
51.9\%           & 55.2\%            & 16.91                       \\ \hline
Cartoon Ratio & Ave Trigger Words  & Ave Description Length \\ \hline
45.6\%              & 36.10           & 104.01                     \\ \hline
\end{tabular}}
 \caption{Numerical properties of the VSD2M dataset.}
\label{numerical_prop}
\end{table}

In order to facilitate the use of researchers, we manually select from multiple pieces of samples corresponding to each top 500 trigger words in real and cartoon domain as test sets, named as VSD-R and VSD-C respectively.
Subsequent experiments on these two test sets are utilized to evaluate the model performance on different domains.

\vspace{-10px}
\subsection{Data Statistics}

Each sample in VSD2M dataset contains a animated or static sticker, a description, a set of trigger words, OCR results, and domain identification (Cartoon/Real).
In the annotated description, in addition to static content such as ``black veins'', the actions such as ``swaying in the wind'' is also well represented.
The annotated caption has more image details and can cover rich visual semantics.
Table \ref{numerical_prop} in shows some numerical properties of VSD2M, including the proportion of cartoon samples, the average number of frames, $etc$, in which real and cartoon domain both account for nearly half.
To further show the details of the dataset, we also count the distribution of static attributes such as trigger words, caption length, $etc.$ in Appendix A.2.
VSD2M aims to provide rich data to the scholars in intelligent creation, thus facilitating the related research in sticker generation. 

\section{Experiments}

\subsection{Experiment Details}

\noindent
\textbf{Models \& Training.}
During inference, we fix the frame number of animated sticker outputs to 8 (consistent with the low-frame-rate characteristic of ASG task and the spatial resolution to $256 \times 256$, with the number of DDIM sampling steps set to 25. For comparative experiments, we select three representative parameter-efficient fine-tuning (PEFT) methods—Customize-A-Video \cite{ren2024customize}, I2V-Adapter \cite{guo2024i2v}, and SimDA \cite{xing2024simda}—due to their reproducibility and reported performance in video generation. All comparative methods follow their original structural designs, with \textit{I2VGen-XL \cite{zhang2023i2vgen} (1.8B parameters)} as the shared base model to ensure fair initial weight consistency. For example, when implementing SimDA, we integrate its Temporal Adapter and Spatial Adapter into each U-Net block of I2VGen-XL, with only the 24M newly added adapter parameters set to learnable.
For RDTF’s training, we use 16 V100-32GB GPUs with a batch size of 1 per GPU, 50K total training steps, and a learning rate of $10^{-5}$ (full-parameter training, no frozen layers). Detailed network configurations of RDTF (e.g., U-Net block connections, STI layer parameters) and condition mask patterns for multi-task training are provided in Appendix.


\noindent
\textbf{Metrics.}
Three metrics are used to evaluate performance:
\textit{Fréchet Video Distance (FVD) \cite{unterthiner2018towards1}}: Measures the distribution consistency between generated and real clips; lower values, better alignment with real data.
\textit{CLIP Similarity \cite{wu2021godiva}}: Evaluates semantic correlation between generated stickers and text captions; all text in testset are used as candidates; higher values, stronger cross-modal alignment.
\textit{Video Quality Assessment (VQA) \cite{wu2022fast}}: Quantifies perceptual quality (e.g., motion smoothness, detail integrity) based on inter-frame contextual relations; higher values, better visual quality.

\vspace{-10px}
\subsection{Compared to State-of-The-Art Methods}

\begin{table}[!h]
\caption{Quantitative comparison of different methods on I\&T$\rightarrow$V task under constrained resources.
\textbf{Bold} and \underline{underline} indicate the best and the second-best, respectively.}
\resizebox{\linewidth}{!}{
\begin{tabular}{c|ccc|c}
\hline
      \textbf{Indicator}                    &  \textbf{Customize-A-Video}        &  \textbf{I2V-Adapter}           &  \textbf{SimDA}                 &  \textbf{Ours}                 \\ \hline
FVD $\downarrow$                      &  451.83                    &        456.24              &       \underline{448.36}                &     \textbf{443.29}     \\
VQA $\uparrow$                      & \underline{0.479}                     &            0.476          &               0.462        & \textbf{0.504}                      \\
CLIP $\uparrow$ & \underline{0.377} & 0.367 & 0.372 & \textbf{0.393} \\ \hline
\end{tabular}}
\label{sota}
\end{table}

\begin{table}[!h]
    \centering
    \caption{Quantitative comparison of different methods for (left) interpolation and (right) prediction task under constrained resources.
     }    
    \begin{minipage}[!h]{.47\linewidth}
    \centering
    \resizebox{\linewidth}{!}{
\begin{tabular}{c|cc}
\hline
       \textbf{Method}                    &     \textbf{FVD $\downarrow$}        &  \textbf{VQA $\uparrow$}                          \\ \hline
                 I2V-Adapter                          & 213.71                                            &   \underline{0.517}       \\
                 SimDA                          &    \underline{198.49}                                        &   0.509       \\                 
 Ours & \textbf{192.62} & \textbf{0.536}  \\ \hline
\end{tabular}}
    \end{minipage}
    \begin{minipage}[htpb]{0.47\linewidth}
    \centering
    \resizebox{\linewidth}{!}{
\begin{tabular}{c|cc}
\hline
       \textbf{Method}                    &     \textbf{FVD $\downarrow$}        &  \textbf{VQA $\uparrow$}                          \\ \hline
                 I2V-Adapter                          &   79.34                                         &   \underline{0.526}       \\
                 SimDA                          &                                    \underline{74.68}        & 0.521         \\                 
 Ours &  \textbf{67.92} & \textbf{0.539}  \\ \hline
\end{tabular}}
    \end{minipage}
    \vspace{-5px}
    \label{interpolation_and_prediction}
\end{table}

Table \ref{sota} presents the quantitative comparison of all methods on the I\&T$\rightarrow$V task (text-image guided animated sticker generation) under resource-constrained conditions. RDTF achieves state-of-the-art performance across all three core metrics: it attains the lowest Fréchet Video Distance (FVD = 443.29), indicating the closest distribution alignment with real sticker data; the highest Video Quality Assessment (VQA = 0.504), verifying superior perceptual quality (e.g., motion smoothness and fine-grained detail retention); and the top CLIP similarity (0.393), confirming strong semantic consistency between generated stickers and text-image guidance. These results collectively validate the effectiveness of RDTF’s dual-mask data utilization and difficulty-adaptive curriculum learning in enhancing detail preservation, semantic alignment, and cross-modal consistency.

Table \ref{interpolation_and_prediction} further reports the evaluation results on two key ASG sub-tasks: interpolation (IPT, left sub-table) and pre/post prediction (PDT, right sub-table). Notably, I2V-Adapter and SimDA require task-specific retraining to adapt to IPT and PDT, while RDTF uses the same set of pre-trained weights for all tasks—demonstrating superior flexibility. Despite this advantage, RDTF still maintains leading performance in both tasks: it achieves the lowest FVD (192.62 for IPT, 67.92 for PDT) and highest VQA (0.536 for IPT, 0.539 for PDT) compared to competitors. This confirms RDTF’s strong generalization capability across multi-condition guided ASG scenarios, which stems from its ability to fully exploit limited data patterns and optimize learning difficulty dynamically.

For qualitative results, please see Appendix B.1.

\subsection{Ablative Study}

\begin{table}[!t]
\centering
\caption{Module ablation results.
DFGN(3D-Conv) represents DFGN using 3D convolution as the temporal layer.
DDU stands for dual-mask based data utilization strategy and DCL stands for difficulty-adaptive curriculum learning.}
\resizebox{0.8\linewidth}{!}{
\begin{tabular}{c|cc}
\hline
    \textbf{Method (I\&T$\rightarrow$V)}  &  \textbf{FVD $\downarrow$} &  \textbf{VQA $\uparrow$} \\
 \hline
DFGN(3D-Conv) &  492.31   & 0.397    \\
DFGN &  478.47   & 0.435    \\
DFGN (w/ DDU) &  459.21   &   0.476  \\
DFGN (w/ DDU \& w/ DCL) &  \textbf{442.18}   & \textbf{0.502}   \\ \hline
\end{tabular}
}
\vspace{-10px}
\label{ablation}
\end{table}

\noindent
\textbf{Module Effectiveness.}
To verify the necessity and synergistic effect of RDTF’s core modules—Spatial-Temporal Interaction (STI) layer, Dual-mask based Data Utilization (DDU) strategy, and Difficulty-adaptive Curriculum Learning (DCL)—we conduct ablation experiments on the I\&T\(\rightarrow\)V task, with results shown in Table \ref{ablation}. All comparisons are based on the DFGN backbone, ensuring the validity of module-specific gains.
First, replacing the 3D convolution temporal layer with the STI layer (from ``DFGN(3D-Conv)'' to ``DFGN'') significantly improves performance: FVD decreases by 13.84 (from 492.31 to 478.47) and VQA increases by 0.038 (from 0.397 to 0.435). This confirms that the STI layer is more suitable for ASG’s low-frame-rate characteristic—it captures cross-frame semantic interactions without redundant 3D convolution computations, enhancing temporal modeling efficiency.
Second, integrating the DDU strategy (``DFGN (w/ DDU)'') further boosts performance: FVD drops by 19.26 (to 459.21) and VQA rises by 0.041 (to 0.476). The gain stems from DDU’s dual advantages: the condition mask reuses data across multi-tasks to expand sample diversity, while the loss mask enhances long-tail long-frame data density—together enriching the model’s learned motion patterns.
Finally, adding the DCL strategy (``DFGN (w/ DDU \& w/ DCL)'') achieves the optimal results: FVD reaches the minimum of 442.18 (a further 17.03 reduction) and VQA peaks at 0.502 (a 0.026 increase). DCL’s ``easy-to-difficult'' sample sequence stabilizes the training process—avoiding early overfitting to hard samples and ensuring the model converges to a better local optimum.
These results demonstrate that each module contributes incrementally, and their synergy (STI for efficient temporal modeling + DDU for data enhancement + DCL for stable convergence) is the key to RDTF’s superior performance under resource constraints.

\begin{table}[!h]
    \centering
    \caption{(Left) Quantitative comparison when using different curriculum learning strategies.
    W/o CL, LCL, and DCL denote without curriculum learning, monotonous curriculum learning, and difficulty-adaptive curriculum learning, respectively.
    (Right) Demonstration of task weight changes in linear curriculum learning.
    }    
    \begin{minipage}[!h]{.5\linewidth}
    \centering
    \resizebox{\linewidth}{!}{
    \begin{tabular}{c|cc}
    \hline
      \textbf{Method  (I\&T$\rightarrow$V)}    &  \textbf{FVD $\downarrow$} &  \textbf{VQA $\uparrow$} \\
     \hline
    RDTF (w/o CL) & 459.21    & 0.476    \\
    RDTF (w/ LCL) & 451.65    & 0.484    \\
    RDTF (w/ DCL) &  \textbf{442.18}   & \textbf{0.502}   \\ \hline
    \end{tabular}}
    \end{minipage}
    \begin{minipage}[htpb]{0.45\linewidth}
    \centering
\includegraphics[width=0.9\columnwidth]{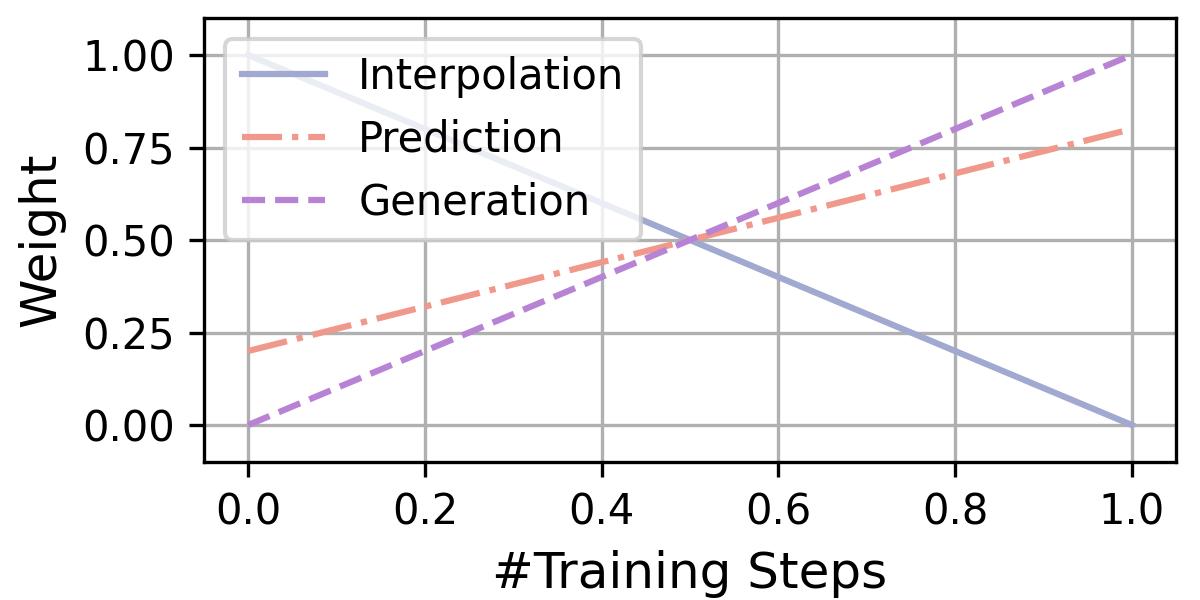}
    \end{minipage}
    \label{CLP}
\end{table}

\noindent
\textbf{Learning Strategy.}
To verify the effectiveness of our proposed Difficulty-adaptive Curriculum Learning (DCL), we compare three strategies: without curriculum learning (w/o CL), linear curriculum learning (LCL), and DCL. The quantitative results are presented in Table \ref{CLP} (left), and the task weight changes in LCL are illustrated in its right sub-figure. As shown in Table \ref{CLP} (left), RDTF (w/o CL) achieves an FVD of 459.21 and a VQA of 0.476. When adopting LCL—where the probability of the interpolation task is gradually reduced and that of prediction/generation tasks is increased linearly—RDTF (w/ LCL) improves to an FVD of 451.65 and a VQA of 0.484. This confirms that learning from easy to difficult tasks benefits model training. Notably, RDTF (w/ DCL) outperforms both, with an FVD of \textbf{442.18} and a VQA of \textbf{0.502}. The improvement stems from DCL’s ability to dynamically adjust sample difficulty via PID-based loss analysis, ensuring stable monotonic entropy growth. In contrast, LCL’s fixed linear task weight changes (right sub-figure of Table \ref{CLP}) cannot adapt to local loss fluctuations, leading to less stable convergence. Thus, DCL’s adaptive mechanism is critical for achieving superior performance under resource-constrained ASG training.

\begin{figure}[!h]
    \centering
    \caption{(Left) Loss comparison during training between RDTF and SimDA.
    (Right) Loss comparison when using linear (LCL) and difficulty-adaptive (DCL) curriculum learning strategy to train RDTF.
    }    
    \begin{minipage}[htpb]{0.46\linewidth}
    \centering
\includegraphics[width=0.99\columnwidth]{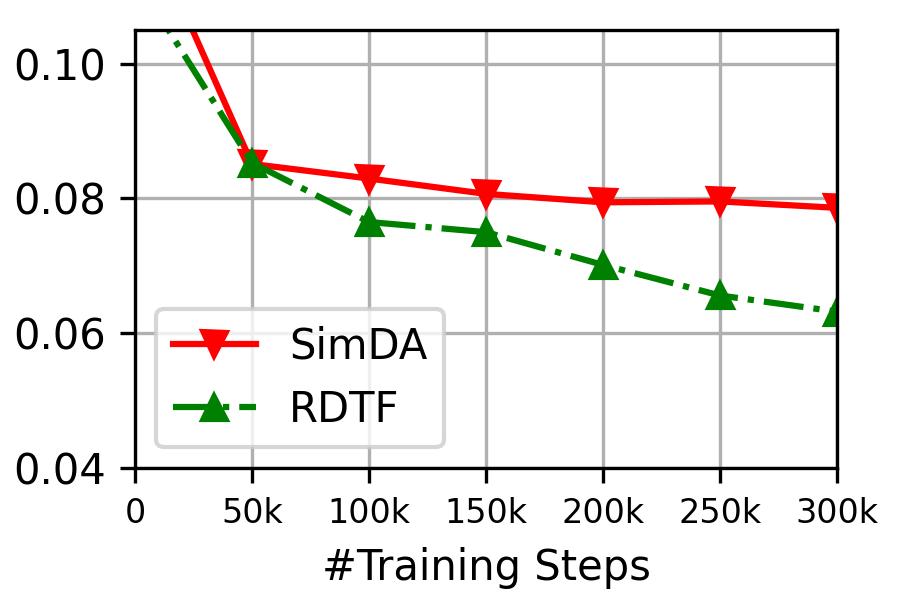}
    \end{minipage}
    \begin{minipage}[htpb]{0.49\linewidth}
    \centering
\includegraphics[width=0.9\columnwidth]{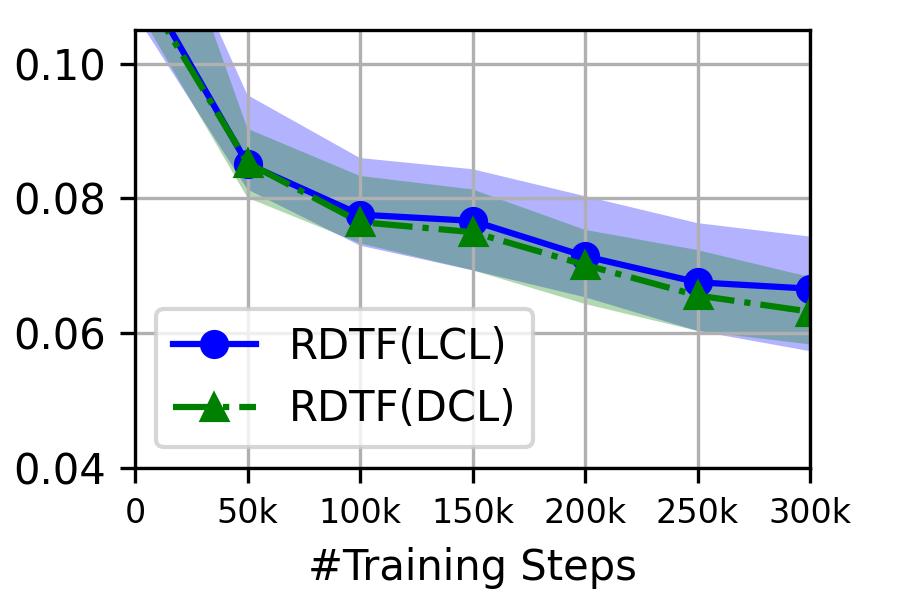}
    \end{minipage}
    \vspace{-5px}
    \label{loss}
\end{figure}

Figure \ref{loss} (left) compares the training loss of RDTF and SimDA. SimDA, with only 24M trainable parameters (vs. RDTF’s full-parameter training), exhibits slower loss decay after 150k steps—indicating limited fitting capability to the animated sticker generation (ASG) target domain. In contrast, RDTF’s loss continues to decrease steadily, demonstrating its advantage in learning task-specific patterns under resource-constrained training.
Figure \ref{loss} (right) further analyzes the impact of curriculum learning strategies. RDTF trained with DCL achieves a smoother and lower loss curve compared to linear curriculum learning. This is because DCL dynamically adjusts sample difficulty based on historical loss statistics, avoiding unstable fluctuations caused by LCL’s fixed linear task weighting—thus ensuring more stable convergence.

\subsection{Perceptual User Study}
\vspace{-5px}

\begin{table}[!h]
\centering
\caption{User study results. Higher scores indicate better user preference.}
\resizebox{0.8\linewidth}{!}{
\begin{tabular}{c|ccc}
\hline
    \textbf{Method  (I\&T$\rightarrow$V)}   &  \textbf{Text.}  &  \textbf{Interest.} &  \textbf{Motion.} \\
 \hline
Customize-A-Video  & \textbf{2.10}   & 1.87 & 1.94  \\
I2V-Adapter &  1.90   & 1.73 & 1.71  \\
SimDA &  1.97   & \underline{2.06}  & \underline{2.03}  \\
RDTF &  \underline{2.03}   & \textbf{2.34}  & \textbf{2.31} \\ \hline
\end{tabular}
}
\label{user_study}
\end{table}

To evaluate the perceptual quality of generated animated stickers from a user perspective, we conduct a perceptual user study with 10 participants (all active users of social communication platforms, familiar with animated sticker scenarios). Participants are asked to rank stickers generated by different methods across three dimensions: \textbf{text alignment} (consistency with input text/image guidance), \textbf{interest} (degree of visual appeal and creativity), and \textbf{motion smoothness} (naturalness of inter-frame transitions). We use \textit{Average User Ranking (AUR)} as the evaluation metric, where a higher score indicates stronger user preference. For text alignment assessment, participants are provided with the corresponding reference text and image to ensure evaluation accuracy.

Table \ref{user_study} presents the AUR results. In \textbf{text alignment}, Customize-A-Video achieves the highest score (2.10), with RDTF closely following (2.03). This slight gap may be attributed to RDTF’s diverse generation results, where a minority of samples exhibit subtle text-detail deviations. However, RDTF dominates in \textbf{interest} (2.34) and \textbf{motion smoothness} (2.31), demonstrating that its generated stickers are more visually engaging and feature more natural motion transitions, which is aligned with the quantitative findings in Table \ref{sota} and visual comparisons in Appendix B.1.


\vspace{-5px}
\subsection{Generalization Performance of the Method}

To verify the generalization of proposed DDU and DCL, we conduct cross-task experiments on two distinct video generation scenarios: medical video generation and autonomous driving video generation. Notably, we strip off the Discrete Frame Generation Network (DFGN) (a module customized for animated sticker generation) and only retain the universal DDU and DCL modules, ensuring the verified generalization stems from the method’s core design rather than task-specific architectures. The experimental results are presented in Table \ref{label1} and Table \ref{label2}.

\begin{table}[!h]
\centering
\caption{Results of different models on the OphNet2024 dataset \cite{hu2024ophnet} for the medical video generation.}
\resizebox{\linewidth}{!}{
\begin{tabular}{c|ccc|c}
\hline
Indicator & Customize-A-Video & I2V-Adapter & SimDA & Ours \\ \hline
FVD $\downarrow$      &       296.2            &      304.7       &    \underline{291.7}   &   \textbf{273.1}   \\
VQA $\uparrow$      &   0.584                &     0.571        &   \underline{0.596}    &  \textbf{0.612}  \\ \hline
\end{tabular}
}
\label{label1}
\end{table}

\textbf{Medical Video Generation.}
We select the OphNet2024 dataset \cite{hu2024ophnet} (focused on ophthalmic medical videos) for this task, as medical videos feature subtle inter-frame changes (e.g., minor movements of ocular structures) and high requirements for distribution consistency—similar to the detail-preserving demand of animated stickers. We preprocess the dataset into 8-frame clips, resulting in 1.1M training samples and 2k validation samples. Since the dataset lacks text annotations, we only evaluate FVD and VQA metrics.

As shown in Table \ref{label1}, our method achieves the lowest FVD (273.1) and highest VQA (0.612) among all competitors. The minimal FVD confirms that the generated medical videos are most consistent with the real data distribution, which is critical for medical scenarios requiring reliability. The leading VQA indicates the generated videos have clear structural details and smooth frame transitions, meeting the basic requirements for clinical auxiliary reference. This gain benefits from DDU’s ability to enhance the information density of limited medical data and DCL’s role in stabilizing the training of high-precision demand tasks.

\begin{table}[!t]
\centering
\caption{Results of different models on the unScene dataset \cite{wen2024panacea} for the autonomous driving video generation.}
\resizebox{\linewidth}{!}{
\begin{tabular}{c|ccc|c}
\hline
Indicator & Customize-A-Video & I2V-Adapter & SimDA & Ours \\ \hline
FVD $\downarrow$      &       320.8           &        311.9     &    \underline{305.6}    &   \textbf{300.4}   \\
VQA $\uparrow$      &    0.351               &           0.347  &    \underline{0.360}   &  \textbf{0.364}  \\ \hline
\end{tabular}
}
\label{label2}
\end{table}

\textbf{Autonomous Driving Video Generation.}
We conduct experiments on the unScene dataset \cite{wen2024panacea} (dedicated to autonomous driving scene videos), which is characterized by dynamic traffic elements and a relatively small data scale. We preprocess it into 8-frame clips, obtaining ~120k training samples and 1k validation samples—consistent with the preprocessing standard of the medical video task.

Table \ref{label2} shows that our method still maintains optimal performance, with the lowest FVD (300.4) and highest VQA (0.364). It is worth noting that SimDA achieves results close to ours (FVD=305.6, VQA=0.360). This is because SimDA, as a parameter-efficient tuning method, has advantages in small-data scenarios due to its lightweight structure. However, our method still outperforms it by virtue of DDU’s data augmentation capability and DCL’s stable convergence mechanism—proving that our core strategies can effectively make up for the limitations of small data volume in autonomous driving video generation.

Overall, the superior performance across medical and autonomous driving video generation tasks fully verifies the strong generalization of our DDU and DCL core strategies, demonstrating their potential for broader low-frame-rate video generation scenarios beyond animated stickers.

\begin{table}[!h]
\centering
 \caption{Performance comparison of training models under different training sets.
}
\resizebox{0.9\linewidth}{!}{
\begin{tabular}{c|cc|cc}
\hline
\multirow{2}{*}{Training Sets} &  \multicolumn{2}{c}{VSD-R} & \multicolumn{2}{c}{VSD-C}   \\
& VQA $\uparrow$ & FVD $\downarrow$ & VQA $\uparrow$ & FVD $\downarrow$ \\
\hline
TGIF \cite{li2016tgif}  & 0.394    &           3310.25                                            &    -                        & - \\
SER30K \cite{liu2022ser30k}  &      -                      &                     -    &    0.372  &             5813.47                             \\
VSD2M    &                                   \textbf{0.487} & \textbf{2613.41}   &    \textbf{0.511}           &                      \textbf{5513.64}        \\ \hline
\end{tabular}
}
\vspace{-15px}
\label{tgif}
\end{table}

\vspace{-5px}
\subsection{VSD2M vs Other Sticker Datasets} 
To verify the strengths of VSD2M compared with other datasets, we use our method to train or fine-tune on different datasets and test on VSD-R and VSD-C testsets.
As shown in Table \ref{tgif}, the VQA and FVD indicators of the model trained on VSD2M have been greatly improved compared with others such as TGIF\cite{li2016tgif} and SER30K\cite{liu2022ser30k}.
This experiment verifies that VSD2M has the ability to generate the well stickers compared with other datasets.

\vspace{-7px}
\section{Conclusion}

This paper focuses on multi-frame animated sticker generation under resource constraints. To address the limitations of parameter-efficient fine-tuning methods in ASG, we propose RDTF—a compact task-specialized framework integrating dual-mask data utilization and difficulty-adaptive curriculum learning. We also construct VSD2M, a million-level multi-modal animated sticker dataset with rich annotations to support RDTF's training from scratch. Experiments show RDTF achieves state-of-the-art results on ASG tasks, outperforming PEFT methods like I2V-Adapter and SimDA quantitatively. This work verifies that small-parameter video generation models, paired with high-quality datasets and optimized training strategies, can outperform PEFT-tuned large models under resource constraints, providing a new paradigm for resource-limited downstream video generation tasks.












\bibliographystyle{unsrt}
\bibliography{sample-base}



\end{document}